\documentclass[floatfix,aps,12pt]{revtex4}
\usepackage{natbib}
\usepackage{amssymb,amsmath}
\usepackage{graphicx}     % Include figure files
\DeclareGraphicsExtensions{.eps}
\graphicspath{{./}}

\newcommand{\mylab}[1]{\label{#1}}
\begin{document}
\title[Depinning of a drop on a rotating cylinder]{On the depinning of a drop of
  partially wetting liquid on a rotating cylinder}
%
%\author[Uwe Thiele]{U\ls W\ls E\ns T\ls H\ls I\ls E\ls L\ls E}
\author{Uwe Thiele}
\affiliation{%
Department of Mathematical Sciences, 
Loughborough University, Leicestershire LE11 3TU, UK}
%\pubyear{2010}
%\volume{}
%\pagerange{}
%\date{\today}
%\setcounter{page}{1}
%
%
\begin{abstract}
  We discuss the analogy of the behaviour of films and drops of liquid
  on a rotating horizontal cylinder on the one hand and substrates
  with regular one-dimensional wettability patterns on the other
  hand. Based on the similarity between the respective governing
  long-wave equations we show that a drop of partially wetting liquid
  on a rotating cylinder undergoes a depinning transition when the
  rotation speed is increased. The transition occurs via a sniper
  bifurcation as in a recently described scenario for drops depinning
  on heterogeneous substrates. \\[3ex]

  {\large The manuscript was accepted by the {\it J. Fluid Mech.} in October
    2010 and will be
    published in the following months.}
%{\bf 78},  041601 (2008) and can be obtained at
%    \href{http://dx.doi.org/10.1103/PhysRevE.78.041601}{http://dx.doi.org/10.1103/PhysRevE.78.041601}
\end{abstract}
\maketitle
%%%%%%%%%%%%%%%%%%%%%%%%%%%%%%%%%%%%%%%%%%%%%%%%%%%%%%%%%%%%%%%%%%%%%%%%%%%%%%%
\section{Introduction} \mylab{sec:intro}
%%%%%%%%%%%%%%%%%%%%%%%%%%%%%%%%%%%%%%%%%%%%%%%%%%%%%%%%%%%%%%%%%%%%%%%%%%%%%%%
%
Motivated by the question how much honey can be kept on a breakfast
knife by rotating the knife about its long axis, Moffatt studied the
destabilisation of a film on the outside of a rotating horizontal
cylinder and the subsequent development of regular patterns of
azimuthal rings \cite[]{Moff77}. He also gives a lubrication equation
that incorporates gravity and viscosity effects but neglects surface
tension and inertia.  It is based on a model by \cite{Pukh77} which
includes surface tension.  Related questions had been studied before
by \cite{YiKi60} employing a rotating horizontal cylinder which is at
the bottom immersed in a liquid bath. i.e., in contrast to
\cite{Moff77} the amount of liquid on the cylinder is not an
independent control parameter.  The related case of a liquid film
covering the inner wall of a horizontal rotating cylinder was first
studied by \cite{Phil60} who discussed experiments performed with
water employing an inviscid model incorporating gravitational and
centrifugal forces.

Since the early experimental results engineers and scientists alike
have studied flows of free-surface films of liquid on the inner or
outer wall of resting or rotating horizontal cylinders in a number of
different settings.  The studied systems are of high importance for
several coating and printing processes where rotating cylinders
transport the coating material in the form of liquid films. In the
present contribution we will discuss the analogy between film flow and
drop motion on a rotating horizontal cylinder on the one hand and on
heterogeneous substrates with regular wettability patterns on the
other hand. First, however, we will give a short account of the
seemingly disconnected fields.

The theoretical analyses given by \cite{YiKi60} and \cite{Phil60} for
the flow on/in a rotating cylinder were based on linear stability
considerations based on the full hydrodynamic equations for
momentum transport, i.e., the Navier-Stokes equations. Their studies
were extended later on, e.g., by \cite{Pedl67} who considered inviscid
film flow on the in- and outside of the cylinder in a unified manner,
and \cite{RuSc76} whose analysis includes viscous film flow at small
and large Reynolds numbers.

However, most analyses of the non-linear behaviour are based on
a long-wave or lubrication approximation \cite[]{ODB97} valid for the
case where the thickness of the liquid film $h(\theta, t)$ is small as
compared to the radius of the cylinder $R$. The resulting evolution
equation for the film thickness profile
\begin{equation}
\partial_t\,h\,=\,-\partial_\theta\, \left\{ h^3\,\left[ 
\tilde{\alpha}\partial_\theta\,\left(\partial_{\theta\theta} h +h \right) 
- \tilde{\gamma} \cos(\theta)\right] + h\right\}
\mylab{eq:film-puk}
\end{equation}
was first given by \cite{Pukh77} and used by \cite{Moff77} (without
capillarity effects, i.e., $\tilde{\alpha}=0$).  In the dimensionless
Eq.~(\ref{eq:film-puk}), $\tilde{\alpha}=3\varepsilon^3\sigma/ R
\omega \tilde{\eta}$ and $\tilde{\gamma}=\varepsilon^2 \rho g
R/3\omega \tilde{\eta}$ are the scaled dimensionless surface-tension
and gravity parameter, respectively \cite[]{HiKe03}.  It is important
to note that both depend on the angular velocity of the rotation
$\omega$, the dynamic viscosity $\tilde{\eta}$, the cylinder radius
$R$, and the ratio $\varepsilon$ of mean film thickness $\bar{h}$ and
$R$.  The remaining material constants are surface tension $\sigma$
and density $\rho$, and $g$ is the gravitational acceleration. The
angle $\theta$ determines the position on the cylinder surface and is
measured anti-clockwise starting at the horizontal position on the
right.
The force $\tilde{\gamma}\cos(\theta)$ corresponds to the component of
gravity parallel to the cylinder surface.

The long-wave equation (\ref{eq:film-puk}) including surface tension
effects has been studied in detail by, e.g., \cite{HiKe03,Pukh05} and
\cite{Kara07}.  Note that for the parameter ranges where
Eq.~(\ref{eq:film-puk}) applies, it is valid without difference for
films on the outside and inside of the cylinder (cf., e.g.,
\cite{HiKe03,AHS03,NKR06}).  A preliminary comparison of the long-wave
approach to results obtained with the Stokes equation is given by
\cite{PJK01}.  Various extensions of Eq.~(\ref{eq:film-puk}) were
developed.  See, in particular, \cite{AcJi04} for a discussion of
hydrostatic effects. Other higher order terms related to gravity,
inertial and centrifugal effects have been included
\cite[]{HoMa99,OBri02,AHS03,ESR04,BeOB05,ESR05,NKR06,Kelm09}.  In
particular, \cite{NKR06} and \cite{Kelm09} present a systematic
derivation of several such higher order models and give also detailed
comparisons to a number of previous studies as e.g., \cite{OBri02} ,
\cite{AHS03} and \cite{BeOB05}.
The film stability and non-linear evolution of a film on the outside
of a non-isothermal horizontal cylinder was considered without
\cite[]{ReBa92} and with \cite[]{DuWi09} rotation. Most of the
aforementioned studies consider two-dimensional situations.  The fully
three-dimensionional problem is studied for the cases without
\cite[]{WSE97} and with \cite[]{NKR05,ESR05} rotation.

The second system that we would like to discuss are liquid drops and
films on heterogeneous solid substrates as studied experimentally,
e.g.~by \cite{ScGa85} and \cite{QAD98}.  Although in principle the
heterogeneity may result from substrate topography or chemical
heterogeneities we focus here on the case of a smooth flat substrate
with chemical heterogeneities, i.e., we assume the substrate
wettability depends on position. Horizontal heterogeneous substrates
are often considered in connection with micro-patterning of soft
matter films via thin film dewetting
\cite[]{Rock99,BaDi00,KKS00,BKTB02}. Another application are
free-surface liquid channels in microfluidics \cite[]{GHLL99}.  Often
one employs thin film evolution equations obtained through a long-wave
approximation to study the dynamics of dewetting for an initially flat
film or to analyse steady drop and ridge solutions and their stability
on such patterned substrates (see, e.g., \cite{KaSh03} and
\cite{TBBB03}).  Static structures may also be studied using a
variational approach \cite[]{BrMa96,LeLi00}.

More recently a model system was introduced to study the dynamics of
drops on heterogeneous inclined substrates. The models apply also to
such drops under other driving forces along the substrate
\cite[]{ThKn06}. The corresponding dimensionless evolution equation
for the one-dimensional film thickness profile (i.e., describing a
two-dimensional drop) is of the form
\begin{equation}
\partial_t\,h\,=\,-\partial_x\, \left\{h^3\,\left[ 
\partial_x\,(\partial_{xx} h + \Pi(h,x))\right]  + \tilde{\mu}\right\},
\mylab{eq:film-depin}
\end{equation}
where $\tilde{\mu}(h)$ represents the lateral driving force ($\mu h^3$
in the case of an inclined substrate, where $\mu$ is the inclination
angle).  The position-dependent disjoining pressure $\Pi(h,x)$ models
the heterogeneous wettability of the substrate. One notices at once
that Eqs.~(\ref{eq:film-depin}) and (\ref{eq:film-puk}) are rather
similar: Identifying $x=\theta$, $\tilde{\mu}(h)=h$ and
$\Pi(h,x)=\tilde{\alpha}h - \tilde{\gamma}\sin(x)$,
Eq.~(\ref{eq:film-puk}) may for $\tilde{\alpha}=1$ be seen as a
special case of Eq.~(\ref{eq:film-depin}). Here, we use the analogy
to study drops or films on a rotating cylinder. In particular, we show
that drops on a rotating cylinder may show rather involved depinning
dynamics analogeous to the behaviour of drops on heterogeneous
substrates \cite[]{ThKn06b,ThKn06,BHT09}.

To illustrate the behaviour we choose the case of drops of a partially
wetting liquid on a substrate with a periodic array of localised
hydrophobic wettability defects \cite[]{ThKn06}. On a horizontal
substrate one finds steady drops that are positioned exactly between
adjacent defects, i.e., the drops stay on the most wettable part. At a
small lateral driving force $\mu$ the $x\to -x$ symmetry of the system
is broken. However, the drops do not slide continuously along the
substrate as they do in the case of a substrate without defects
\cite[]{Thie01}. Their advancing contact line is blocked by the next
hydrophobic defect, i.e., one finds steady pinned drops at an upstream
position close to the defect. As the driving force is increased, the
drops are pressed further against the defect. As a result the steady
drops steepen. At a finite critical value $\mu_c$ the driving force
allows the drop to overcome the adverse wettability gradient and the
drop depins from the defect and begins to slide.  In a large part of
the parameter space spanned by the wettability contrast and the drop
size the point of depinning at $\mu_c$ corresponds to a Saddle-Node
Infinite PERiod (sniper) bifurcation. Beyond $\mu_c$ the drops do not
slide continuously down the incline. They rather perform a stick-slip
motion from defect to defect \cite[]{ThKn06b,ThKn06}. However, in
another region of the parameter space the depinning can occur via a
Hopf bifurcation, i.e., with a finite frequency. In particular, this
happens in the case of rather thick wetting layers.

The aim of the present contribution is to elucidate under which
conditions a similarly intricate `depinning' behaviour can be found
for a pendant drop underneath a horizontal cylinder when increasing
the speed of rotation. In the following section \ref{sec:model} we
will discuss our scaling and introduce a long-wave model for a film of
partially wetting liquid on a rotating cylinder. It allows (even
without gravitation) for the coexistence of drops and a wetting
layer. We discuss several physical situations it can be applied
to. Next, section \ref{sec:res} presents steady state solutions
depending on the non-dimensional rotation velocity for several
thicknesses of the wetting layer and discusses the depinning scenario
in the case of the partially wetting liquid.  Finally, we conclude and
give an outlook in Section~\ref{sec:conc}.

%%%%%%%%%%%%%%%%%%%%%%%%%%%%%%%%%%%%%%%%%%%%%%%%%%%%%%%%%%%%%%%%%%%%%%%%%%%%%%%
\section{Model}
\mylab{sec:model}
%%%%%%%%%%%%%%%%%%%%%%%%%%%%%%%%%%%%%%%%%%%%%%%%%%%%%%%%%%%%%%%%%%%%%%%%%%%%%%%
%
Before we introduce our long-wave model we discuss our scales. In the
non-dimen\-sion\-ali\-sation used in most works studying
Eq.~(\ref{eq:film-puk}) the angular velocity of the cylinder rotation
$\omega$ is not reflected in a single dimensionless parameter but both
dimensionless parameters are inversely proportional to $\omega$.
Furthermore, the time scale depends on $\omega$
\cite[]{Pukh77,HiKe03}. This and a further re-scaling of the equation
for steady drop and film states, that absorbs the flow rate into both
dimensionless parameters and also into the thickness scale, implies
that the existence of two steady solutions for identical parameters
(thin capillary film and pendant drop with through-flow), shown
analytically by \cite{Pukh05} and numerically by \cite{Kara07},
actually refers to solutions of different liquid volume.

Such a scaling is not convenient for our study as we would like to
investigate the change in system behaviour for fixed liquid volume
when increasing the rotation speed of the cylinder from zero, i.e.,
$\omega$ should only enter a single dimensionless parameter that will
be our main control parameter.  A scaling fit for our purpose is the
one based on the scales of the gravity-driven drainage flow employed
by \cite{ESR04}.

Note also that in particular \cite{Kara07} and to a lesser extent
\cite{WSE97} represent some of their solutions in a way that
misrepresents their findings and, in general, the concept of an
equation in long-wave approximation. Whenever one shows the solutions
as a thickness profile on a cylinder one has to pick a particular
radius $R$ \textit{and} smallness ratio $\epsilon$ for the
representation (cf.~Figs.~\ref{fig:rot-steady-prof},
\ref{fig:rot-prof} and \ref{fig:rot-wetting-prof} below). This is a
rather arbitrary choice as the equation in long-wave approximation is
strictly valid only in the limit $\epsilon\to0$, ie. it is always only
an approximation to a real physical situation where $\epsilon$ is
finite.

\begin{figure}
\centering
\includegraphics[width=0.4\hsize]{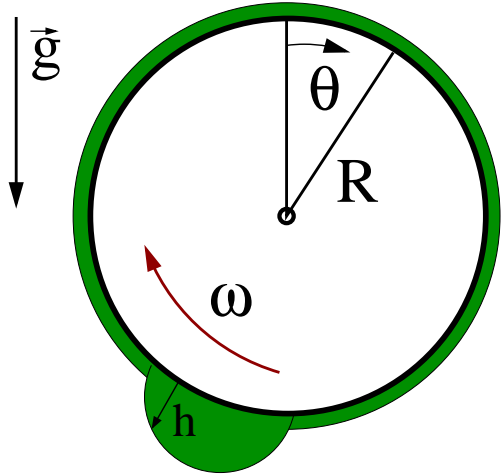}
\caption{Sketch of a drop of partially wetting liquid coexisting with a thick 
wetting layer on a rotating cylinder.
}
\mylab{fig:sketch}
\end{figure}

In the present work we focus on solutions without axial variation,
i.e., we consider the two-dimensional physical situation depicted in
Fig.~\ref{fig:sketch}. Following \cite{ESR04} we use $\bar{h}$, $R$,
$\rho g \bar{h}^2/\eta$, $\varepsilon\rho g \bar{h}^2/\eta$,
$3\eta/\varepsilon\bar{h}\rho g$ and $\rho g \bar{h}$ to scale the
coordinate $z$ orthogonal to the cylinder surface, the coordinate
$x=\theta R$ along the surface, the two velocity components $v_x$ and
$v_z$, time, and pressure, respectively. The ratio of $z-$ and
$x-$scale is the smallness parameter $\varepsilon=\bar{h}/R$, i.e.,
the film height is scaled by $\bar{h}=\varepsilon R$.  Here the angle
$\theta$ is defined in clockwise direction starting at the upper
vertical position, i.e., a pendant droplet underneath the resting
horizontal cylinder has its centre of mass at $\theta=\pi$.

The evolution equation is derived from the Navier-Stokes equations
with no-slip boundary conditions at the (rotating) cylinder, and
tangent- and normal force equilibria at the free surface using the
long-wave approximation \cite[]{Pukh77,ReBa92,ODB97,ESR04}. Note that
here we furthermore assume that the ratio $\bar{h}/R$ and the physical
equilibrium contact angle are both of $O(\varepsilon)$
(cf.~Section~\ref{sec:conc}).  This gives
\begin{equation}
\partial_\tau\,h\,=\,-\partial_\theta\, \left\{h^3\,
  \partial_\theta\,\left[\mathrm{Bo}^{-1}\left(\partial_{\theta\theta} h + h\right) 
    - \cos(\theta) + \widetilde{\Pi}(h)\right]   + \widetilde{\Omega} h
\right\},
\mylab{eq:film}
\end{equation}
where $\tau$ is the non-dimensional time, and
\begin{equation}
\mathrm{Bo}=\frac{R^3\rho g}{\bar{h}\sigma}
\quad\mathrm{and}\quad
\widetilde{\Omega}=\frac{\eta\omega R}{\rho\bar{h}^2 g}\ge 0
\mylab{eq:bond}
\end{equation}
are an effective Bond number and a rotation number for a clock-wise
rotation, respectively.  The latter corresponds to the ratio of the
rotation velocity at the cylinder surface and the drainage
velocity. The function $\widetilde{\Pi}(h)$ is a non-dimensional
disjoining pressure that accounts for the wettability of the liquid
film on the cylinder surface \cite[]{deGe85,Isra92,Bonn09}.

To make the analogy with the depinning droplet on a heterogeneous
substrate more obvious we introduce another timescale
$t=\tau/\mathrm{Bo}$, disjoining pressure $\Pi=\mathrm{Bo}\widetilde{\Pi}$
and rotation number $\Omega=\mathrm{Bo}\widetilde{\Omega}$ and obtain
\begin{equation}
\partial_t\,h\,=\,-\partial_\theta\, \left\{h^3\,
  \partial_\theta\,\left[\partial_{\theta\theta} h + h
    - \mathrm{Bo}\cos(\theta) + \Pi(h)\right]  + \Omega h
\right\},
\mylab{eq:film-sca}
\end{equation}
with
\begin{equation}
\Omega=\frac{\eta\omega R}{\varepsilon^3\sigma}.
\mylab{eq:bond}
\end{equation}
Defining the derivative
\begin{equation}
\partial_h f = -\Pi(h) - h + \mathrm{Bo}\cos(\theta)
\mylab{eq:fh}
\end{equation}
of a `local free energy' $f(h,\theta)$ and identifying
$\tilde{\mu}=\Omega h$, Eq.~(\ref{eq:film-sca}) takes the form of
Eq.~(\ref{eq:film-depin}) and the Bond number takes the role of a
`heterogeneity strength'. Note that the scaling used is well adapted
for discussing independently the influences of rotation and gravity in
the presence of capillarity. However, it does not allow for a study of the
limit of vanishing surface tension.

As disjoining pressure we choose here the combination of a long- and a
short-range power law $\Pi(h)=\mathrm{Ha}/h^3\,(1 - b/h^3)$, where Ha
is a dimensionless Hamacker constant. For $b>0$ one has a precursor
film or wetting layer thickness $h_0=b^{1/3}$
\cite[]{Pism01,PiTh06,Thie10}. This is valid only for Ha$<0$, i.e.,
when the long-range interaction is destabilising. This corresponds to
the case of a partially wetting liquid.  For Ha$>0$, a rather thick
wetting layer can be stabilised and the $b/h^6$ term is of no further
relevance and could just as well be dropped.

Note that $\partial_h f$ may contain other terms beside the ones in
Eq.~(\ref{eq:fh}). For a heated or cooled cylinder a term
$(3/2)\,\mathrm{Bo}\mathrm{Bi}\,\mathrm{Ma}\,\log [h/(1 +
\mathrm{Bi}\,h)] + 1/(1 + \mathrm{Bi}\,h)]$ would enter where Bi is
the Biot number and Ma an effective Marangoni number
\cite[]{ReBa92,OrRo92,ThKn04}.  If the rotating cylinder forms the
inner electrode of a cylindrical capacitor, a term
$\mathrm{Vo}/[h+(d-h)\epsilon]^2$ has to be included for a dielectric
liquid and DC voltage of non-dimensional strength Vo ($\epsilon$ is
the electric permittivity and $d$ the distance between the two
cylinders (see appendix of \cite{ThKn06} and \cite{JHT08}).  The
similarity in the behaviour of depinning drops under the influence of
different physical effects in the case that the effective pressure
terms `look similar' is discussed by \cite{ThKn06}.

Although here we do not consider such other physical effects
explicitly, we will explore the effect of corresponding thick wetting
layers by adapting the parameters of our disjoining pressure
accordingly. A thick wetting layer that coexists with droplets might,
for instance, result from the interplay of a destabilising thermal or
electrical effect and a stabilising long-range van der Waals
interaction.  We obtain adequate values of Ha and $b$ for our present
$\Pi(h)$ by relating them to the (non-dimensional) thickness of the
wetting layer $h_0$ and a static macroscopic contact angle $\beta_0$
by \cite[see, e.g.,][]{deGe85}
\begin{equation}
b=h_0^3\qquad\mbox{and}\qquad\mathrm{Ha}=-\frac{5}{3}\beta_0^2h_0^2.
\end{equation}
Note, that $\beta_0$ is the angle in long-wave scaling, i.e., a small
physical equilibrium contact angle
$\beta_\mathrm{eq}=\varepsilon\beta_0$ corresponds to a long-wave
contact angle $\beta_0$ of $O(1)$.
In the limit of a non-rotating cylinder (i.e.\ $\Omega=0$) and
assuming the disjoining pressure corresponds to complete wetting
(Ha$>0$, $1/h^6$ term may be dropped), our equation
(\ref{eq:film-sca}) corresponds to the one given by \cite{ReBa92} for
the isothermal case.  There, however, first a viscous scaling is used,
which they transform in a second step -- a rescaling of time and
therefore velocities with the Galileo number -- into the `drainage
scaling' used here.
Our equation corresponds also to that of \cite{HiKe03} when the
disjoining pressure is added to their equation.

To analyse the system behaviour we determine in the following
steady-state solutions that correspond, e.g., to pendant droplets. The
behaviour beyond the depinning threshold is analysed using
time-stepping algorithms. As we restrict ourselves to the
two-dimensional physical situation an explicit scheme for stiff
equations suffices for the latter. The steady-state solutions are
obtained using the continuation techniques of the package AUTO
\cite[]{AUTO2000}. The steady and time-periodic solutions are
characterized by their $L^2$ norm $||\delta h||\equiv\sqrt{(1
  /2\pi)\int_0^{2\pi}(h(\theta)-1)^2d\theta }$ and time-averaged $L^2$
norm\\ $||\delta h||\equiv\sqrt{(1 /2\pi
  T)\int_0^T\int_0^{2\pi}(h(\theta)-1)^2d\theta dt}$,
respectively. $T$ is the time period.

Note finally that Eq.~(\ref{eq:film-sca}) is the result of our choices
for the relative order of magnitude of the involved dimensionless
numbers that is based on the choice of physical effects that shall be
discussed (cf.~\cite{ODB97}). Once this is accepted, equation
(\ref{eq:film-sca}) is to $O(1)$ asymptotically correct. Its form as a
conservation law implies that $\int_0^{2\pi} hd\theta=2\pi$. To $O(1)$
this corresponds to mass conservation.  For a discussion of higher
order corrections to this picture see \cite{Kelm09}.

%%%%%%%%%%%%%%%%%%%%%%%%%%%%%%%%%%%%%%%%%%%%%%%%%%%%%%%%%%%%%%%%%%%%%%%%%%%%%%%
\section{Results}
\mylab{sec:res}
%%%%%%%%%%%%%%%%%%%%%%%%%%%%%%%%%%%%%%%%%%%%%%%%%%%%%%%%%%%%%%%%%%%%%%%%%%%%%%%
\subsection{Parameters}
\mylab{sec:param}

We base our estimate for realistic Bond and rotation numbers on two
liquids typically studied in the literature: (i) water at
$25^{\circ}$C as in \cite{ReBa92} and a silicone oil with
$\sigma=0.021$N/m, $\eta=1$kg/ms and $\rho=1200$kg/m$^3$ as in
\cite{ESR04}. As experimentally feasible configurations we assume
angular rotation velocities $\omega=0.1\dots10$s$^{-1}$, cylinder of
radii $R=10^{-3} \dots10^{-2}$m and smallness ratios
$\varepsilon=0.01\dots0.1$. It is only for the purpose of illustration
of steady droplets on a cylinder that we later use $\varepsilon=0.1$
in selected figures.

For $\omega=1$s$^{-1}$ and a smallness ratio $\varepsilon=0.1$ we find
for a cylinder of radius $R=10^{-2}$m for material (i) Bo$=136$ and
$\Omega=0.14$, whereas material (ii) gives Bo$=561$ and
$\Omega=476$. For a smaller cylinder radius of $R=10^{-3}$m one has
(i) Bo$=1.36$ and $\Omega=0.014$, and (ii) Bo$=5.61$ and
$\Omega=47.6$. Decreasing the smallness ratio to $\varepsilon=10^{-2}$
increases all Bond numbers by a factor 10 and all rotation numbers by
a factor $10^3$. Focusing on cylinder diameters in the millimetre
range, we mainly investigate Bo$\le10$.  Note that experiments may
also be performed with larger cylinders. To keep the Bond number in
the interesting range one could decrease the relevant density by
replacing the ambient gas by a second (immiscible) liquid. A feasible
experimental set-up could be a horizontal Taylor-Couette apparatus
with rotating inner cylinder. The theoretical framework would also
need to be amended -- replacing the one-layer theory
[Eq.~(\ref{eq:film-sca})] by a closed two-layer model similar to
\cite{MPBT05}.

Note, that silicone oils may have viscosities 2 orders of magnitude
smaller or larger than the chosen value and one is able to vary the
velocity of rotation in a wide range. This implies that there is some
flexibility in the choice of $\Omega$. We will find that for Bo$=O(1)$
the interesting range for the rotation number is $\Omega=O(1)$. For
instance, with Bo$=1$ one finds that depinning occurs at
$\Omega=1.68$.

As the present work aims at establishing the qualitative
correspondence between the behaviour of a depinning drop on a
heterogeneous substrate and the one of a droplet on a rotating
cylinder we restrict our results to one partially wetting case
(choosing $h_0=0.1$ and $\beta_0=2$) and the completely wetting case
($\beta_0=0$). For comparison, selected results are also shown for the
contact angle $\beta_0=1$.

Note that here the scaling and therefore the discussion of the
smallness parameter $\epsilon$ is based on $\epsilon=\bar{h}/R$, i.e.,
we use the ratio of mean film thickness and cylinder radius. In the
case of pendant drop solutions one also has to keep the maximal film
height $h_\mathrm{max}$ in mind when discussing $\epsilon$ as the
scaling becomes questionable when $h_\mathrm{max}\gg\bar{h}$. This is,
however, normally not the case. In the present work the ratio
$h_\mathrm{max}/\bar{h}$ is always below 3.  Although an inspection of
the figures in \cite{Kara07} shows a ratio $h_\mathrm{max}/\bar{h}$
that is not much larger, one notices that there $\epsilon$ itself
(based on $\bar{h}$) seems to be larger than one (see remark above
Eq.~(\ref{sec:model})). The same applies to Fig.~3 of \cite{WSE97}.
There exists no such problem in the region where only a thin wetting
layer of thickness $h_\mathrm{min}$ covers the cylinder even if
$h_\mathrm{min}\ll\bar{h}$. Actually, the long-wave approximation gets
better there. In an approach based on matched asymptotics one could
further simplify the governing equation in this region
(cf.~\cite{BBK08}). However, such an approach is not taken here as it
is of limited use when considering parameter regions that show
qualitative changes in solution behaviour, e.g., close to the sniper
bifurcation discussed below.

\subsection{Partially wetting case}
\mylab{sec:part-wet}

\begin{figure}
\centering
\includegraphics[width=0.7\hsize]{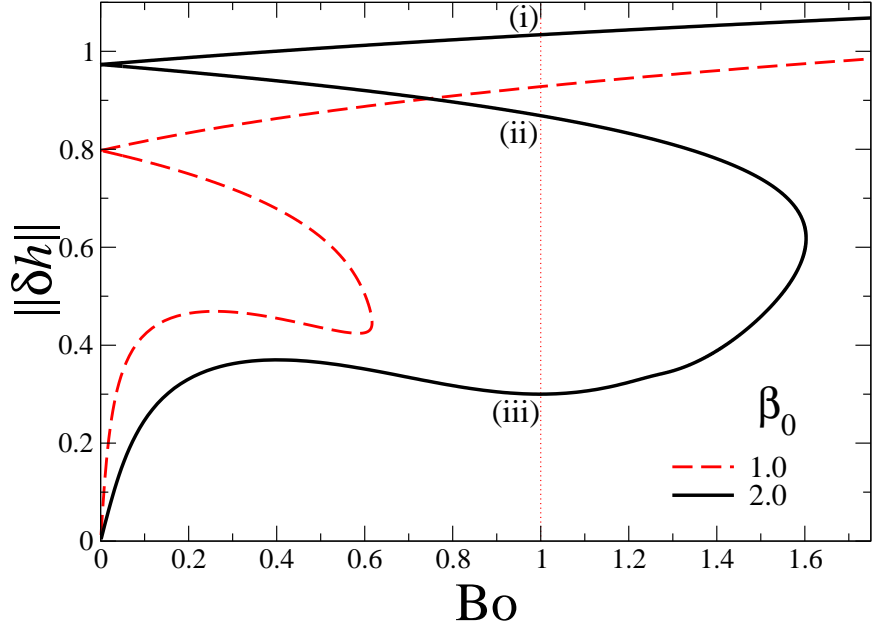}
  \caption{Solutions on a horizontal cylinder without rotation ($\Omega=0$) 
are characterised by their $L^2$ norm as a function of the Bond number
for two different equilibrium contact angles $\beta_0$ and
$h_0=0.1$, $\bar{h}=1.0$. 
}
\mylab{fig:rot-steady-bif}
\end{figure}

\begin{figure}
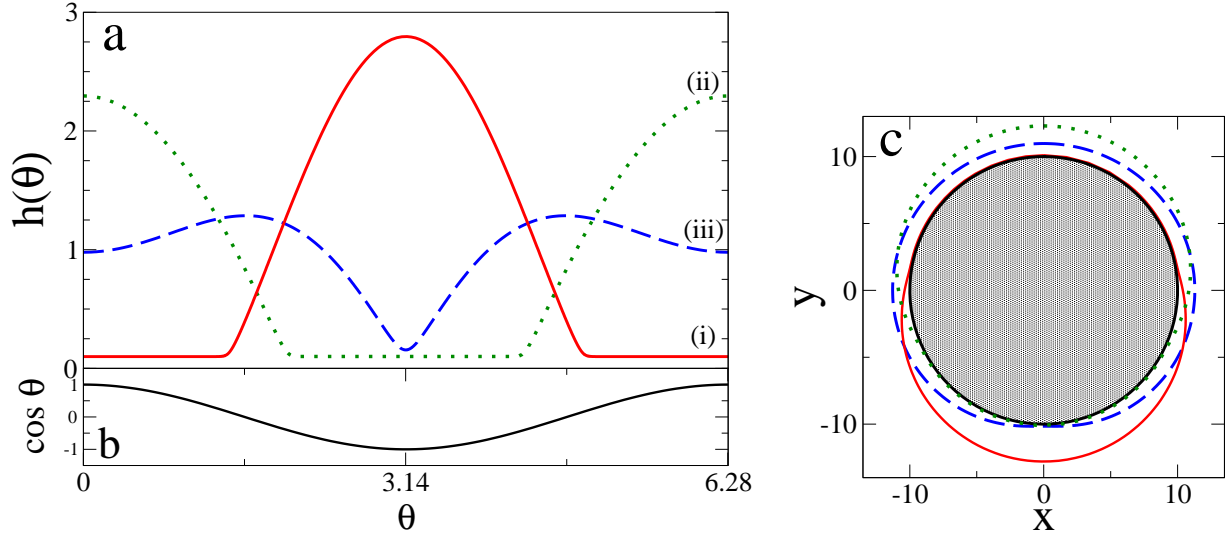

\centering
\includegraphics[width=0.6\hsize]{prof_bo1,0per6,28eps0vel0,0hmean1hp0,1theta2,0}\hspace{0.025\hsize}
\includegraphics[width=0.35\hsize]{prof_bo1,0per6,28eps0vel0,0hmean1hp0,1theta2,0_polarcoord_R10}
\caption{ Steady film thickness profiles on a horizontal cylinder
  without rotation ($\Omega=0$) for Bo$=1$ and $\beta_0=2.0$
  (corresponding to marks (i) to (iii) in
  Fig.~\ref{fig:rot-steady-bif}).  The remaining parameters are as in
  Fig.~\ref{fig:rot-steady-bif}. Panel (a) gives the profiles as
  $h(\theta)$, (b) gives the space-dependent part of $\partial_h f$
  (Eq.~(\ref{eq:fh})), and panel (c) shows the profiles on a
  cylinder. Note that for illustration purposes we have (rather
  arbitrarily) assumed a radius $R=10$, i.e. $\varepsilon=0.1$
  (cf.~discussion in Section~\ref{sec:param}).  Line styles in (c)
  correspond to those in (a). The cylinder surface is represented by
  the solid black line.}
  \mylab{fig:rot-steady-prof}
\end{figure}

First, we consider the case of a horizontal cylinder without rotation
($\Omega=0$) and determine steady drop and film solutions. Employing
the Bond number as control parameter we obtain families of steady
state solutions for two selected values of the equilibrium contact
angle $\beta_0$. Inspecting Fig.~\ref{fig:rot-steady-bif} one notices
that above a critical value of the Bond number Bo$_\mathrm{c}$ there
exists only a single solution. It corresponds to a symmetric pendant
drop located underneath the cylinder. An example for Bo$=1$ and
$\beta_0=2.0$ is given as solid line in
Fig.~\ref{fig:rot-steady-prof}.  However, two additional steady
solutions exist below Bo$_\mathrm{c}$. One of them corresponds to an
unstable symmetric drop sitting on top of the cylinder (dotted line in
Fig.~\ref{fig:rot-steady-prof}). The other one corresponds to an
unstable solution that has two minima -- a deep one underneath the
cylinder and a shallow one on top of it (dashed line in
Fig.~\ref{fig:rot-steady-prof}).

We find that Bo$_\mathrm{c}$ becomes smaller with decreasing contact
angle $\beta_0$, i.e., the range of Bo numbers where steady
structures beside the pendant drop exist becomes smaller. The
completely wetting case ($\beta_0=0$) is qualitatively different and
is discussed below in Section~\ref{sec:comp-wet}.  Increasing
$\beta_0$ above $\beta_0=2$ the critical value Bo$_\mathrm{c}$
strongly increases, e.g., for $\beta_0=10$ one finds
Bo$_\mathrm{c}=43.3$.

Note, that the pendant drop underneath the cylinder [drop on top of cylinder]
corresponds on a horizontal heterogeneous substrate to the drop on the
more [less] wettable region as discussed by \cite{TBBB03}. The limit
of zero Bond number is analogue to the horizontal homogeneous
substrate.

\begin{figure}
\centering
\includegraphics[width=0.7\hsize]{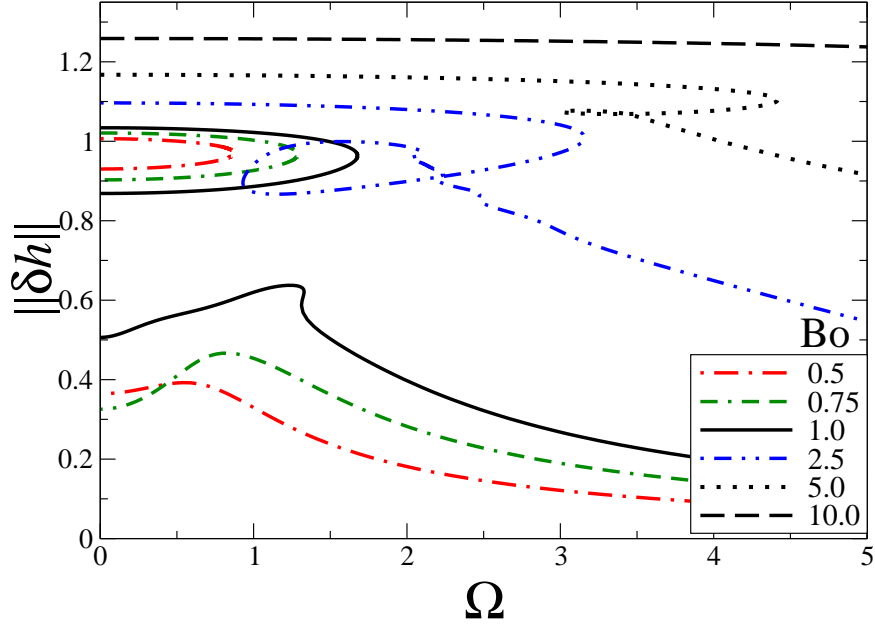}
\caption{Solutions on a horizontal rotating cylinder as a function of
  the Rotation number $\Omega$ for various Bond number Bo as given in
  the legend and contact angle $\beta_0=2.0$. The drop and film
  profiles are characterised by their $L^2$ norm.  The remaining
  parameters are $h_0=0.1$, and $\bar{h}=1.0$.  }
\mylab{fig:rot-bif}
\end{figure}

\begin{figure}
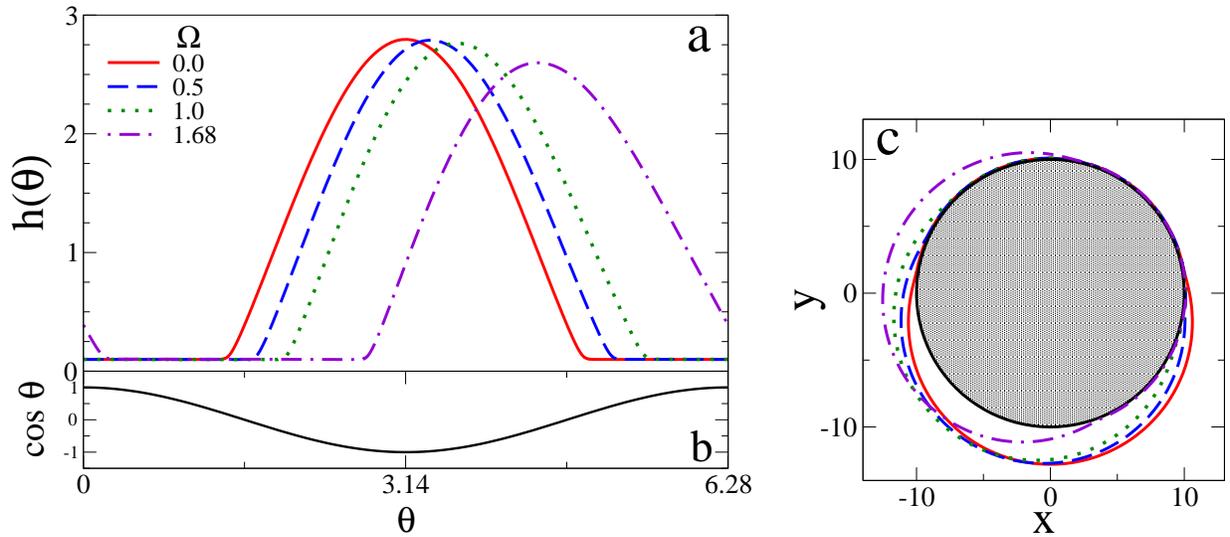

\centering
\includegraphics[width=0.6\hsize]{prof_diffvel_bo1,0per6,28eps0hmean1hp0,1theta2,0}\hspace{0.025\hsize}
\includegraphics[width=0.35\hsize]{prof_diffvel_bo1,0per6,28eps0hmean1hp0,1theta2,0_polarcoord_R10}
\caption{ Steady film thickness profiles on a rotating horizontal cylinder
  for Bond number Bo$=1$ and
  $\beta_0=2.0$ for Rotation numbers $\Omega$ as given in the legend. 
All shown solutions are situated on the stable upper branch in 
  Fig.~\ref{fig:rot-bif}.  The remaining parameters are as in
  Fig.~\ref{fig:rot-bif}. Panel (a) gives the profiles as
  $h(\theta)$, (b) gives the space-dependent part of $\partial_h f$
  (Eq.~(\ref{eq:fh})), and panel (c) illustrates the profiles on a
  cylinder assuming a radius $R=10$.  Line styles in (c) correspond to
  those in (a).}
\mylab{fig:rot-prof}
\end{figure}

Next, we increase the dimensionless angular velocity $\Omega$ for
several selected Bond numbers and determine the steady state solutions
(Fig.~\ref{fig:rot-bif}). Focusing first on Bo$=1$ we recognise at
$\Omega=0$ three solutions as marked by the vertical dotted line in
Fig.~\ref{fig:rot-steady-bif}. The curve of largest norm corresponds
to the pendant drop. As $\Omega$ increases from zero, its norm and
amplitude decrease slightly, whereas its centre of mass moves towards
larger $\theta$, it is `dragged along' by the rotation without
changing its shape much. This can be well appreciated in
Fig.~\ref{fig:rot-prof} which gives selected profiles for the branch
of pendant drop solutions.  The branch terminates in a saddle-node
bifurcation at $\Omega_\mathrm{sn}=1.68$ where it annihilates with one
of the unstable branches.  The third branch, i.e., the one of lowest
norm, shows for increasing $\Omega$ an increase in the norm, then it
undergoes two saddle-node bifurcations at $\Omega=1.313$ and
$\Omega=1.329$. At larger $\Omega$ the norm decreases again.  Note
that the two saddle-node bifurcations are a first sign of rather
complicated re-connections which occur when one increases Bo from
$1.0$ to $2.5$. The re-connections involve further (unstable) branches
that are not shown in Fig.~\ref{fig:rot-bif} and result in the loop
structure observed for Bo$=2.5$ at about $\Omega=1.5$.  A further
indication for the existence of additional branches is the observation
that the value of the norm at $\Omega=0$ for Bo$=1.0$ in
Fig.~\ref{fig:rot-bif} does not agree with the lowest norm at Bo$=1.0$
in Fig.~\ref{fig:rot-steady-bif}. As the re-connections are not
relevant for the depinning transition we will here not consider them
further.

\begin{figure}
\centering
\includegraphics[width=0.7\hsize]{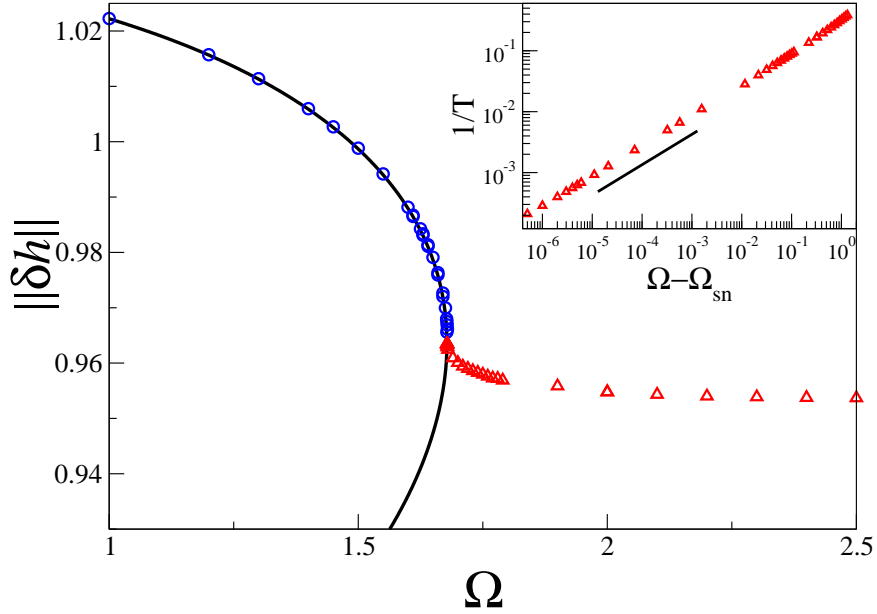}
\caption{Bifurcation diagram for the depinning transition for a drop
  on a rotating cylinder with Bo$=1$ and $\beta_0=2.0$. Shown are the
  $L^2$ norm for the branch of steady solutions as obtained by
  continuation (solid line), selected steady solutions as obtained by
  direct integration in time (circles) and the time-averaged $L^2$
  norm for the unsteady solutions beyond the sniper depinning
  bifurcation (triangles). The inset gives for the latter a log-log
  plot of the dependence of the inverse temporal period on the
  distance from the bifurcation $\Omega-\Omega_\mathrm{sn}$.  The solid line
  indicates an exponent of $1/2$.
  \mylab{fig:bifrot} }
\end{figure}

\begin{figure}
\centering
\hspace*{-1cm}(a)\includegraphics[width=0.49\hsize]{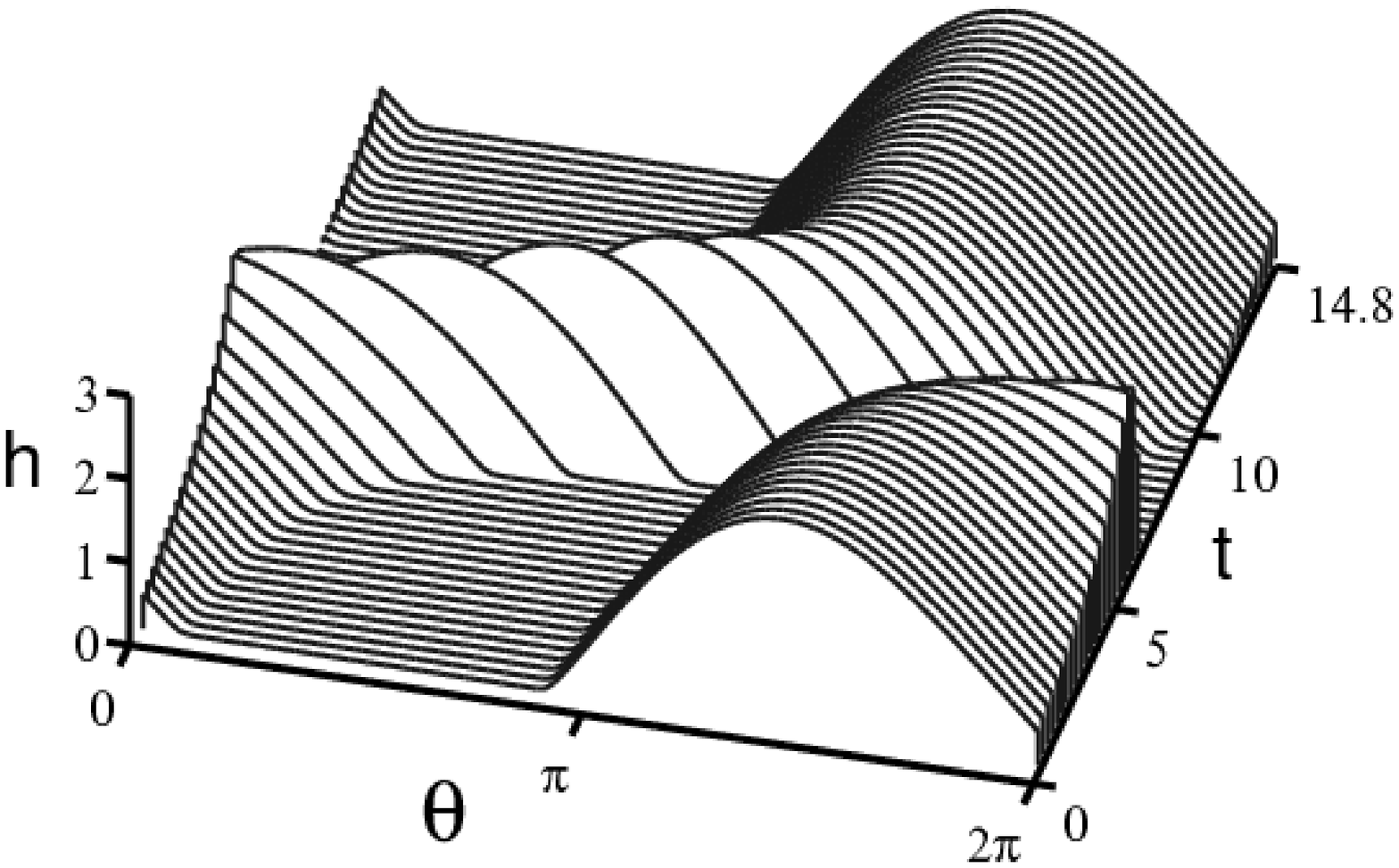}
(b)\includegraphics[width=0.49\hsize]{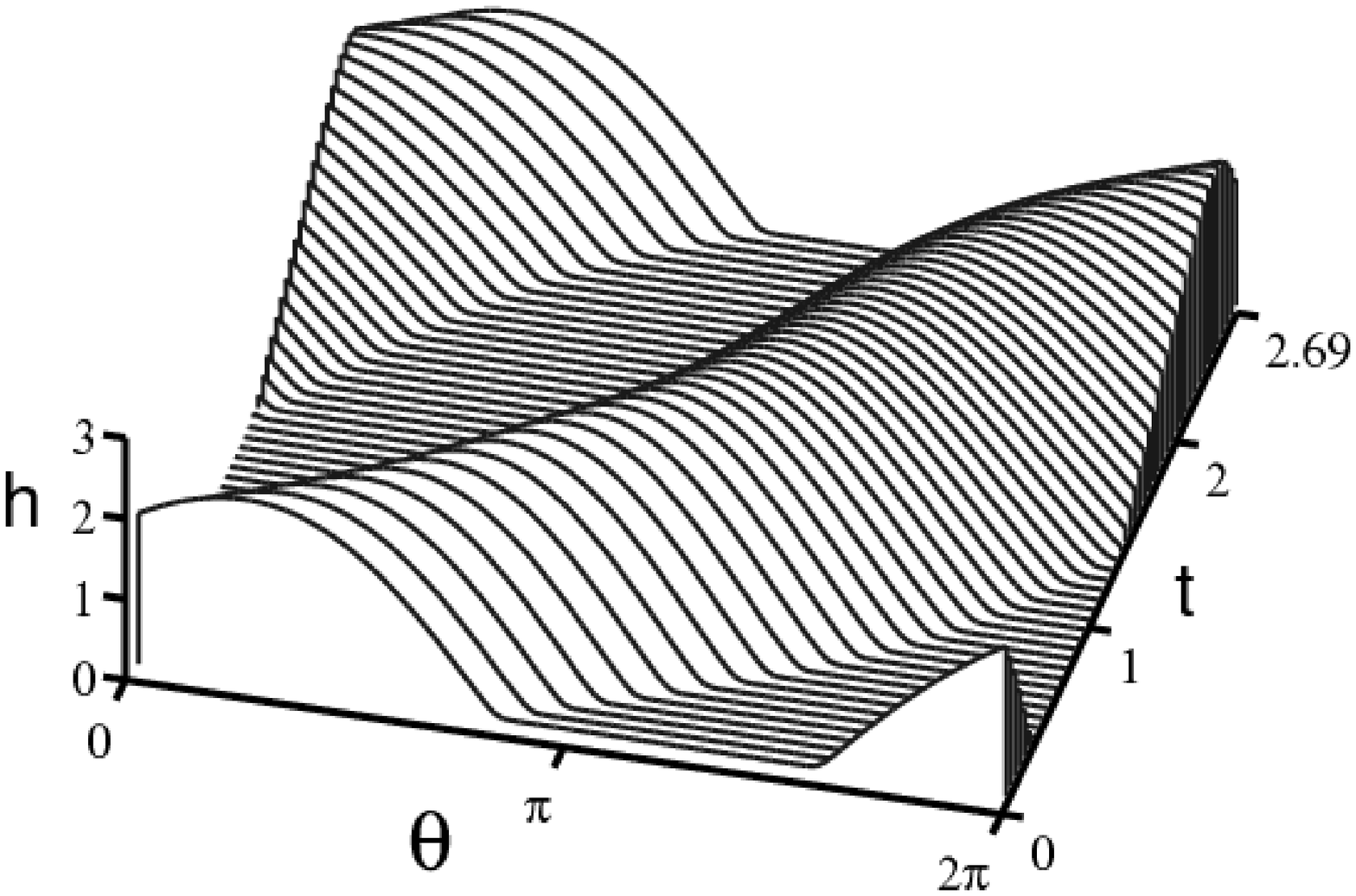}
\caption{Space-time plots illustrating the time evolution of
  co-rotating drops beyond depinning via a sniper bifurcation (at
  $\Omega>\Omega_\mathrm{sn}$). Shown is one period in space and time
  (a) close to depinning at Rotation number $\Omega=1.8$ with a
  temporal period of $T=14.8$, and (b) far from depinning at
  $\Omega=3.0$ with $T=2.7$.  The Bond number is Bo$=1$ and
  $\beta_0=2.0$.  }
\mylab{fig:rot-depin-stplot}
\end{figure}

At Bond numbers smaller than Bo$=1$ the behaviour is qualitatively the
same as described for Bo$=1$. However, the critical
$\Omega_\mathrm{sn}$ becomes smaller with decreasing Bo.  Above, we
have discussed that the behaviour at $\Omega=0$ changes qualitatively
when decreasing Bo beyond Bo$_c$
(Fig.~\ref{fig:rot-steady-bif}). Although this also affects the
behaviour at small $\Omega$ it has less influence on the behaviour at
larger $\Omega$ (Fig.~\ref{fig:rot-bif}).  In particular, the
saddle-node at $\Omega_\mathrm{sn}$ where the pendant drop solution
ceases to exist, persists for larger Bond number as does the related
dynamic behaviour (see below).  Note that for Bo$=10$ the saddle-node
is at $\Omega_\mathrm{sn}\approx6.2$ outside the range of
Fig.~\ref{fig:rot-bif}.

An interesting resulting question is what happens to a stable
obliquely pendant droplet at $\Omega<\Omega_\mathrm{sn}$ when the
rotation velocity is further increased such that it is slightly larger than
$\Omega_\mathrm{sn}$? Does the solution approach the remaining steady
solution?
Simulations of Eq.~(\ref{eq:film-sca}) show that this is not the
case. For $\Omega>\Omega_\mathrm{sn}$ the droplet moves in a
non-stationary way with the rotating cylinder and corresponds to a
space- and time-periodic solution. The resulting bifurcation
diagram for Bo$=1$ is given in Fig.~\ref{fig:bifrot}. It shows that
the `new' branch of space- and time-periodic solutions (characterised
by its time-averaged norm) emerges from the saddle
node bifurcation of the steady solutions at $\Omega_\mathrm{sn}$.
The behaviour close to and far from the saddle-node bifurcation is
illustrated in the space-time plots of
Fig.~\ref{fig:rot-depin-stplot}. Panel (b) shows that far away from
the bifurcation the drop moves continuously with its velocity and
shape varying smoothly. It moves fastest when its maximum passes
$\theta\approx\pi/2$ (velocity $\approx 4.6$) and slows down when
passing $\theta\approx3\pi/2$ (velocity $\approx 1.2$), i.e.,
respectively, when gravity most strongly supports and hinders the
motion driven by the cylinder rotation. The ratio of largest to
smallest velocity is about 4:1.  When approaching the bifurcation from
above, the droplet still moves with the rotation but the time scales
of the slow and the fast phase become very different. For instance, at
$\Omega=1.8$ (Fig.~\ref{fig:rot-depin-stplot}~(a)) the drop moves
fastest when passing $\theta\approx\pi/2$ (velocity $\approx
3.3$). However, when the drop is situated at about $\theta=3\pi/2$ it
barely moves (velocity $\approx 0.1$) resulting in a velocity ratio of
33:1 between the fastest and slowest phase. The resulting overall
behaviour strongly resembles the stick-slip motion discussed in the
context of contact line motion on heterogeneous substrates
\cite[]{ThKn06}. Note that it also resembles so-called sloshing modes
found for a partially liquid-filled rotating cylinder
\cite[]{ThMa97,HoMa99}.

The ratio of the velocities diverges when approaching
the bifurcation point.  In consequence, the frequency $1/T$ of the
periodic drop motion goes to zero. As shown in the inset of
Fig.~\ref{fig:bifrot}, the dependence corresponds to a power law
$1/T\sim(\Omega-\Omega_\mathrm{sn})^{1/2}$. This indicates that the
bifurcation at $\Omega_\mathrm{sn}$ is actually a
Saddle-Node-Infinite-PERiod bifurcation (or `sniper' for short; see
\cite{Stro94}, and discussion by \cite{ThKn06}).

The described behaviour is quite generic for the partially wetting
case, i.e., it is found in a wide range of parameters, in particular,
Bond number, contact angle and wetting layer thickness.  
However, the behaviour changes dramatically when strongly 
decreasing the contact angle, i.e., when approaching the 
completely wetting case. 

\subsection{The completely wetting case}
\mylab{sec:comp-wet}

After having discussed the intricate depinning behaviour in the
partially wetting case, to emphasise the contrast, we consider briefly
the completely wetting case ($\beta_0=0$).  Note that the physical
setting is then identical to the one employed in most studies of the
classic Moffatt problem \cite[]{Moff77,Pukh77}.  However, the
parametrisation used here is different and results in Eq.~(\ref{eq:film-sca})
with Ha$=0$. We analyse the system behaviour as above by studying the
steady state solutions without and with rotation.

\begin{figure}
\centering
\includegraphics[width=0.7\hsize]{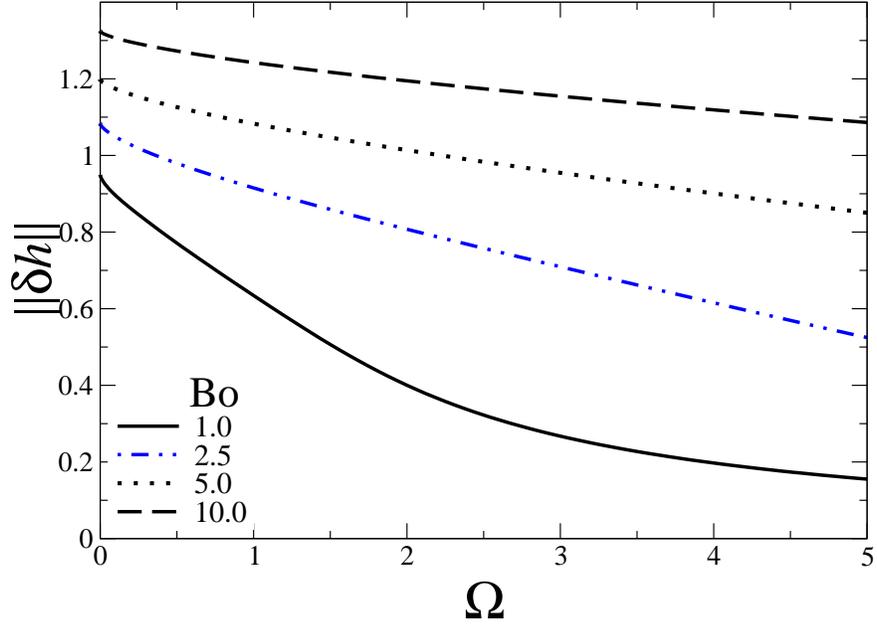}
\caption{Solutions for a film of completely wetting fluid
  ($\beta_0=0$) on a horizontal rotating cylinder as a function of
  the Rotation number $\Omega$ for various Bond numbers Bo as given in
  the legend. The solutions are characterised by their $L^2$ norm. The
  remaining parameter is $\bar{h}=1.0$.}
\mylab{fig:rot-wetting}
\end{figure}                     

\begin{figure}
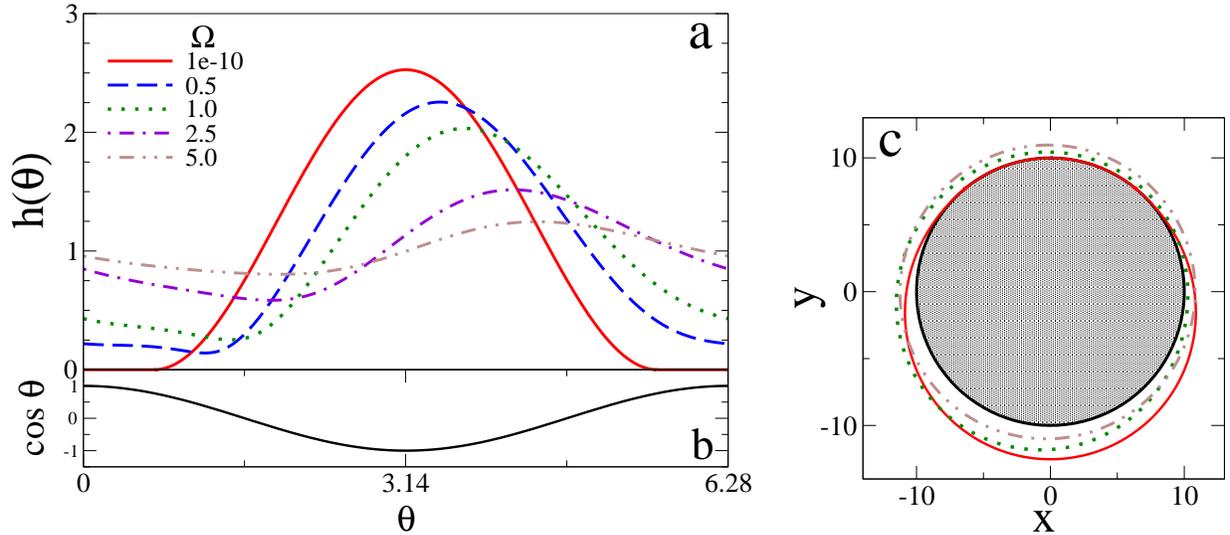

\centering
\includegraphics[width=0.6\hsize]{prof_diffvel_bo1,0per6,28eps0hmean1hp0,1theta0,0}\hspace{0.025\hsize}
\includegraphics[width=0.35\hsize]{prof_diffvel_bo1,0per6,28eps0hmean1hp0,1theta0,0_polarcoord_R10}
\caption{Steady film thickness profiles for the case of complete wetting
  with Bo$=1$ for Rotation numbers $\Omega$ as given in the legend. 
The remaining parameters are as in
  Fig.~\ref{fig:rot-wetting}. Panel (a) gives the profiles as
  $h(\theta)$, (b) gives the space-dependent part of $\partial_h f$
  (Eq.~(\ref{eq:fh})), and panel (c) illustrates the profiles on a
  cylinder assuming a radius $R=10$.  Line styles in (c) correspond to
  those in (a).}
\mylab{fig:rot-wetting-prof}
\end{figure}

In the case without rotation, the first difference one notices is that
for fixed Bond number only one solution exists. It corresponds to the
pendant drop solution. With increasing Bo, i.e., with increasing
importance of gravity, its amplitude monotonically increases. Fixing
Bo and increasing the rotation velocity from zero, one finds a
monotonic decrease of the norm (Fig.~\ref{fig:rot-wetting}) and height
of the drop. Its maximum moves towards larger angular position
$\theta$, i.e., the liquid is dragged along with the
rotation. However, in the wetting case there is no force besides
gravity that can favour drops as compared to a flat film. As gravity
acts downwards it only favours pendant drops.  Therefore, the drop
does not survive as a `coherent structure' when dragged upwards with
the rotation: it is flattened and smeared out. Typical profiles are
given in Fig.~\ref{fig:rot-wetting-prof}. Note that we do not show the
exact case without rotation, but choose $\Omega=10^{-10}$ as then
a dynamically created wetting layer still exists that is numerically
advantageous (cf.~\cite[]{ThKn04}).

When increasing $\Omega$ further than shown in
Fig.~\ref{fig:rot-wetting} one finds that the thickness profiles
approaches a flat film. For instance, for Bo$=1$ the norm decreases
below $10^{-2}$ at $\Omega\approx70$.

%%%%%%%%%%%%%%%%%%%%%%%%%%%%%%%%%%%%%%%%%%%%%%%%%%%%%%%%%%%%%%%%%%%%%%%%%%%%%%%
\section{Conclusion}
\mylab{sec:conc}
%%%%%%%%%%%%%%%%%%%%%%%%%%%%%%%%%%%%%%%%%%%%%%%%%%%%%%%%%%%%%%%%%%%%%%%%%%%%%%%
%
Based on the observation that the equations in long-wave approximation
that govern film flow and drop motion (i) on or in a rotating cylinder
and (ii) on a heterogeneous substrate are rather similar, we have
explored whether the analogy can be exploited, i.e., whether it allows
results obtained for one system to be transferred to the other
one. In particular, we have found that indeed for drops on a rotating
cylinder there exists a counterpart of the rather involved depinning
dynamics described recently for drops on heterogeneous substrates.

To study the effect we have introduced an alternative scaling. This
has been necessary because the commonly used scaling contains the
angular velocity of the cylinder rotation in both dimensionless
parameters and in the time scale.  Together with a further re-scaling
of the steady state equation involving the flow rate
\cite[]{Pukh77,Kara07} this does not allow for an investigation of
either the system behaviour when increasing the rotation speed of the
cylinder from zero or the existence of multiple solutions for fixed
rotation speed and liquid volume. The scaling employed here does allow
for such studies as the rotation speed only enters a single
dimensionless parameter.  Furthermore, it allows us to discuss the
analogy between the two systems of interest in a rather natural
way. In particular, downward gravitation and rotation speed for the
drop on the rotating cylinder correspond to the heterogeneous
wettability and lateral driving force, respectively, for drops on
heterogeneous substrates.

Guided by this analogy we have studied the behaviour of drops of
partially wetting liquid on the rotating cylinder.  As a result it has
been shown, how `switching on' gravity (increasing the Bond number)
without rotation changes the solution behaviour dramatically as the
`gravitational heterogeneity' along the cylinder surface effectively
suppresses multidrop solutions until only the single pendant droplet
underneath the cylinder survives. We have furthermore found that
increasing the rotation from zero to small values, the drop is dragged
along to a stable equilibrium position where downwards gravity and
upwards drag compensate. At a critical speed, however, the retaining
downwards force is too small and the drop is dragged above the left
horizontal position.  It then continuously moves with the rotating
cylinder in a non-constant manner. Close to the transition the drop
shows stick-slip motion. An analysis of the time-dependent behaviour
has shown that the frequency related to the periodic motion goes to
zero at the threshold following a power law with power $1/2$. This and
the related bifurcation diagram show that the observed transition
corresponds to depinning via a sniper bifurcation. The behaviour found
is rather generic for the case of partial wetting. Although, here we
have presented results for particular parameter values, it can be
observed in a wide range of parameters, in particular, Bond number,
equilibrium contact angle and wetting layer thickness.
However, we have also found that the behaviour changes dramatically
when strongly decreasing the equilibrium contact angle, i.e., when
approaching the completely wetting case that is normally studied in
the literature.  When increasing the velocity of rotation in this case
we have seen a transition from pendant drops underneath the cylinder
to a nearly uniform film around the cylinder as \cite{ESR04}.

The presented results indicate that it may be very fruitful to further
explore the analogy of films/drops on rotating cylinders and
heterogeneous substrates. We expect that the recent exploration of the
three-dimensional case for depinning drops \cite[]{BHT09} and for the
various transitions between ridge, drop and rivulet states
\cite[]{BKHT11} based on tools for the continuation of steady and
stationary solutions of the two-dimensional thin film equation
(physically three-dimensional system) \cite[]{BeTh10} is also relevant
for the case of a rotating cylinder. A future study could, e.g.,
elucidate the relation between the formation of azimuthal rings,
pendant ridges and sets of drops.

The present circular cylinder is an analogy to a sinusoidal
wettability profile.  Cylinders with other cross sections, for
instance, an elliptical cylinder \cite[]{Hunt08} would correspond to
more localised defects of the corresponding heterogeneous
substrate. However, the analogue system turns out to be more complicated
as a rotating non-circular cylinder introduces the aspect of a
time-periodic non-harmonic forcing parallel to the substrate into the
heterogeneous substrate system.

Our equation (\ref{eq:film-sca}) is derived under two conditions: (i)
the mean (and also the maximal) film thickness has to be small as
compared to the radius of the cylinder; and (ii) the local surface
slope (e.g., the physical equilibrium contact angle) has to be
small. Here we have assumed that the two related smallness parameters
are of the same order of magnitude.  In a next step one may introduce
two different smallness parameters and discuss a number of distinct
limits in dependence of their ratio.  Such a systematic asymptotic
study would also permit to discuss the influence of other effects on
the depinning behaviour like, for instance, the effects of the
hydrostatic pressure \cite[]{AHS03,AcJi04}, centrifugal forces
\cite[]{ESR04} and inertia \cite[]{HoMa99,Kelm09}.

\section{Acknowledgements}

I acknowledge support by the EU via grant PITN-GA-2008-214919
(MULTIFLOW). First steady state solutions where calculated by
N.~Barranger employing a predecessor of the present model.
I benefited from discussions with E. Knobloch and several group members
at Loughborough University as well as from the input of several
referees and the editor.

\end{document}